\documentclass[twocolumn,pre,showpacs,10pt]{revtex4-1}
\usepackage{graphicx}
\usepackage{amsmath}
\usepackage{amsfonts}
\usepackage[table]{xcolor}
\usepackage[caption=false]{subfig}
\usepackage{array}
\usepackage{floatrow}
\usepackage{pbox}
\usepackage[colorlinks=true,linkcolor=blue,citecolor=blue,filecolor=cyan,urlcolor=blue,breaklinks=true]{hyperref}
\usepackage{psfrag}

\begin{document}
\title{Oscillations in feedback driven systems: thermodynamics and noise}

\author{Daniele De Martino$^{1}$ and Andre C Barato$^{2}$}
\affiliation{$^{1}$Jozef Stefan Institute, Jamova Cesta 39, 1000 Ljubjlana, Slovenia\\
$^2$ Department of Physics, University of Houston, Houston, Texas 77204, USA}

\parskip 1mm
\def\d{{\rm d}}
\def\Ps{{P_{\scriptscriptstyle \hspace{-0.3mm} s}}}
\def\MF{{\mbox{\tiny \rm \hspace{-0.3mm} MF}}}
\def\ts{\tau_{\textrm{sig}}}
\def\tos{\tau_{\textrm{osc}}}
\begin{abstract}
Oscillations in nonequilibrium noisy systems are important physical phenomena. These oscillations 
can happen in autonomous biochemical oscillators such as circadian clocks. They can also manifest as subharmonic oscillations 
in periodically driven systems such as time-crystals. Oscillations in autonomous systems and, 
to a lesser degree, subharmonic oscillations in periodically driven systems have been both 
thoroughly investigated, including their relation with thermodynamic cost and noise.
We perform a systematic study of oscillations in a third class of nonequilibrium systems:
feedback driven systems. In particular, we use the apparatus of stochastic thermodynamics 
to investigate the role of noise and thermodynamic cost in feedback driven oscillations. 
For a simple two-state model that displays oscillations, we analyze the relation between
precision and dissipation, revealing that oscillations can remain coherent for an indefinite
time  in a finite system with thermal fluctuations in a limit of diverging thermodynamic cost.
We consider oscillations in a more complex system with several degrees of freedom, an Ising model driven
by feedback between the magnetization and the external field. This 
feedback driven system can display subharmonic oscillations similar to the ones observed in time-crystals. 
We illustrate the second law for feedback driven systems that display oscillations. For the Ising model, the oscillating dissipated 
heat can be negative. However, when we consider the total entropy that also includes an informational  term related to 
measurements, the oscillating total entropy change is always positive. We also study the finite-size 
scaling of the dissipated heat, providing evidence for the existence of a first-order phase transition
for certain parameter regimes. 
\end{abstract}

\maketitle

\section{Introduction}

Oscillations are a phenomena  of paramount importance in physics, biology, 
chemistry, and economy. They can happen on  scales ranging from 
microscopic to astronomical. They often take place in autonomous nonequilibrium noisy 
systems that dissipate energy to sustain the oscillations.
Prominent examples are autonomous biochemical oscillators, such as 
systems of interacting molecules that display circadian rhythms driven by the consumption of 
chemical energy \cite{vanz07,dong08}.

Fluctuations can fundamentally change the behavior of biochemical oscillations \cite{bark00,gonz02,mein02,falc03,mcka07}. 
For instance, noise can generate oscillations, in the sense that a  biochemical system that has no oscillations in its deterministic description with  nonlinear rate equations 
can display oscillation at the the level of a stochastic  description that accounts for fluctuations in the finite number of chemical species. Oscillations is such finite 
noisy systems also have  a limited precision. In fact, the relation between the precision of biochemical oscillations and the amount of dissipated energy required to maintain 
them, analyzed through the lens of stochastic thermodynamics \cite{seif12}, has been the subject of several works \cite{qian00,cao15,bara17a,nguy18,fei18,wier18,mars19}.
Another example of an autonomous nonequilibrium oscilator recently analyzed with the theory of stochastic thermodynamics, is the so-called electron-shuttle \cite{wach19}. 

A second class of non-equilibrium noisy systems that display oscillations is a certain phase of periodically driven many body systems known as time-crystal. 
Time-crystals  are systems driven by a time-periodic Hamiltonian that display oscillations with a period larger than the period of the drive, so-called 
subharmonic oscillations. Time-crystals have been studied in closed quantum systems \cite{sach15,khem16,else16} and also open systems that dissipate energy
\cite{laza17,gong18,wang18,gamb19}. Besides displaying spontaneous symmetry breaking of time-reversal symmetry, i.e., the onset of subharmonic oscillations, 
they also display spatial long range order. The relation between thermodynamics and the precision of subharmonic oscillations in finite stochastic 
systems has also been investigated with the theory of stochastic thermodynamics in \cite{ober19}.

Hitherto we have mentioned two classes of nonequilibrium systems,  autonomous systems driven by a fixed  thermodynamic force such as biochemical oscillators  and periodically driven systems such as time-crystals. 
A third class of nonequilibrium systems showing oscillations are feedback driven systems \cite{bech05}, which are of central importance in engineering and technology \cite{astr10}.  
These systems are driven  out of equilibrium by measurement and feedback, i.e., a change in the Hamiltonian of the system that depends 
on the measurement outcome. Within stochastic thermodynamics, feedback driven systems have played an important role in elucidating the relation between information and thermodynamics \cite{cao09,saga12,parr15}. 
In particular, the total entropy for feedback driven systems includes an informational term related to the increase in entropy generated by the controller that performs the  measurements and applies the feedback.  
An important general feature of deterministic feedback driven systems is that they start to develop oscillations when the controller tries to fix the system onto unstable states, 
in the presence of non-linearities \cite{astr10,andr66}. A basic understanding of the thermodynamics of these oscillatory phenomena is still lacking.   

In this paper, we provide a systematic analysis of the thermodynamics of temporal oscillations in stochastic feedback driven systems. 
We start with a simple two-state model that displays oscillations and that can be solved exactly. We then proceed to study a more complex system with several degrees of freedom,
a fully interacting Ising model driven by a feedback scheme, which is discrete in time, between the external field and the magnetization. 
This is a generalization of  the model with continuous feedback introduced in \cite{dema19}, which has been shown to display self-oscillations that persist in the thermodynamic limit below 
the static critical temperature.

The two-state model introduced here provides a paradigmatic, exactly solvable, example of oscillations in feedback driven systems. 
We show that the number of coherent oscillations, an observable that quantifies the precision of noisy oscillations, can be arbitrarily large in finite feedback driven systems subjected to thermal fluctuations. 
The size of the system does not impose a fundamental bound on the precision of oscillations in feedback driven systems, in contrast to autonomous systems, for which the number of 
coherent oscillations is fundamentally bounded  by the number of states \cite{bara17a}. We also analyze the relation between thermodynamic cost and precision for this two-state system.

We show that feedback driven systems display a phase similar to time-crystals: the Ising model with discrete feedback we introduce here displays 
subharmonic oscillations with a period larger than the time-interval between two measurements, which can be taken as the natural period of a feedback 
driven systems. The oscillations in the magnetization, 
which  persist indefinetely in the thermodynamic limit, take place for temperatures below the critical  static temperature.

Concerning the scaling of the rate of entropy production per spin with system size, we show that in the thermodynamic limit this rate 
is zero above the critical temperature and is larger than zero below the critical temperature. At criticality, the rate of entropy production per spin can either go to 
zero with a mean field exponent or it can be finite, which correspond to second-order and first-order phase transitions, respectively. 

Thermodynamic quantities such as heat and work also oscillate below 
the critical temperature. We show that while the oscillating dissipated heat can be negative, if we also include the informational contribution to the total entropy change that appears in the second law 
for feedback driven systems, the total entropy change is positive for all times.

The paper is organized in the following way. In Sec. \ref{sec2} we analyze the two-state model. Sec. \ref{sec3} is dedicated to the 
Ising model. We conclude in \ref{sec4}. Appendix \ref{appa} contains a brief introduction to the stochastic thermodynamics of feedback driven systems. The finite-size scaling 
analysis of the rate of entropy production for the Ising model with continuous feedback is reported in Appendix \ref{appb}.


\section{Two-state model with feedback}
\label{sec2}
\subsection{Definition of the model}

We first introduce a simple feedback driven system that displays oscillations. A general 
definition of thermodynamic quantities in feedback driven systems is provided in Appendix \ref{appa}. The system consists of
a single spin with  two states $s=\pm 1$. The energy is given by $E_s=-hs$, where $h$ 
is the external magnetic field. The spin is in contact with a heat bath at temperature 
$T$, it flips between these two states due to thermal fluctuations for a time-interval $\tau$.
We assume that the dynamics of the system during this time-interval is Markovian. The master equation for 
the evolution of the probability to be at state $s=1$ during the nth time-interval reads
\begin{equation}
\frac{d}{dt}p_n(t)= w^{h_n}_1[1-p_n(t)]-w^{h_n}_2p_n(t),
\label{eqmaster}
\end{equation}
where $w^{h_n}_1$ is the transition rate from state $s=-1$ to state $s=1$ and $w^{h_n}_2$ 
is the reversed transition rate. These transition rates fulfill the detailed balance condition
\begin{equation}
\frac{w_2^{h_n}}{w_1^{h_n}}= \textrm{e}^{-2\beta h_n},
\label{eqDB}
\end{equation}
where $\beta=1/(k_BT)$ and $k_B$ is the Boltzmann constant that is set to $k_B=1$ throughout.

A feedback driven system is also characterized by measurement and feedback. 
At the end of a time-interval, the state of the system is measured without measurement error. 
The feedback scheme is such that if the system at the end of 
the nth time-interval is in the state $s_n=1$ ($s_n=-1$),  then the magnetic field for the next time-interval
is set to $h_{n+1}=-h_0$ ($h_{n+1}=h_0$), where $h_0\ge 0$. We note that the transition rates in Eq. \eqref{eqmaster}
are not fixed quantities but rather they depend on the state of the system in the previous time-interval, i.e., 
they are random quantities that  depend on the particular stochastic trajectory. 

\subsection{Oscillatory behavior}

This feedback scheme generates oscillations on the average spin orientation at the end of 
a time-interval as a function of $n$. In the following we show this property with the exact 
calculation of the average spin orientation.

\begin{figure}
\includegraphics[width=77mm]{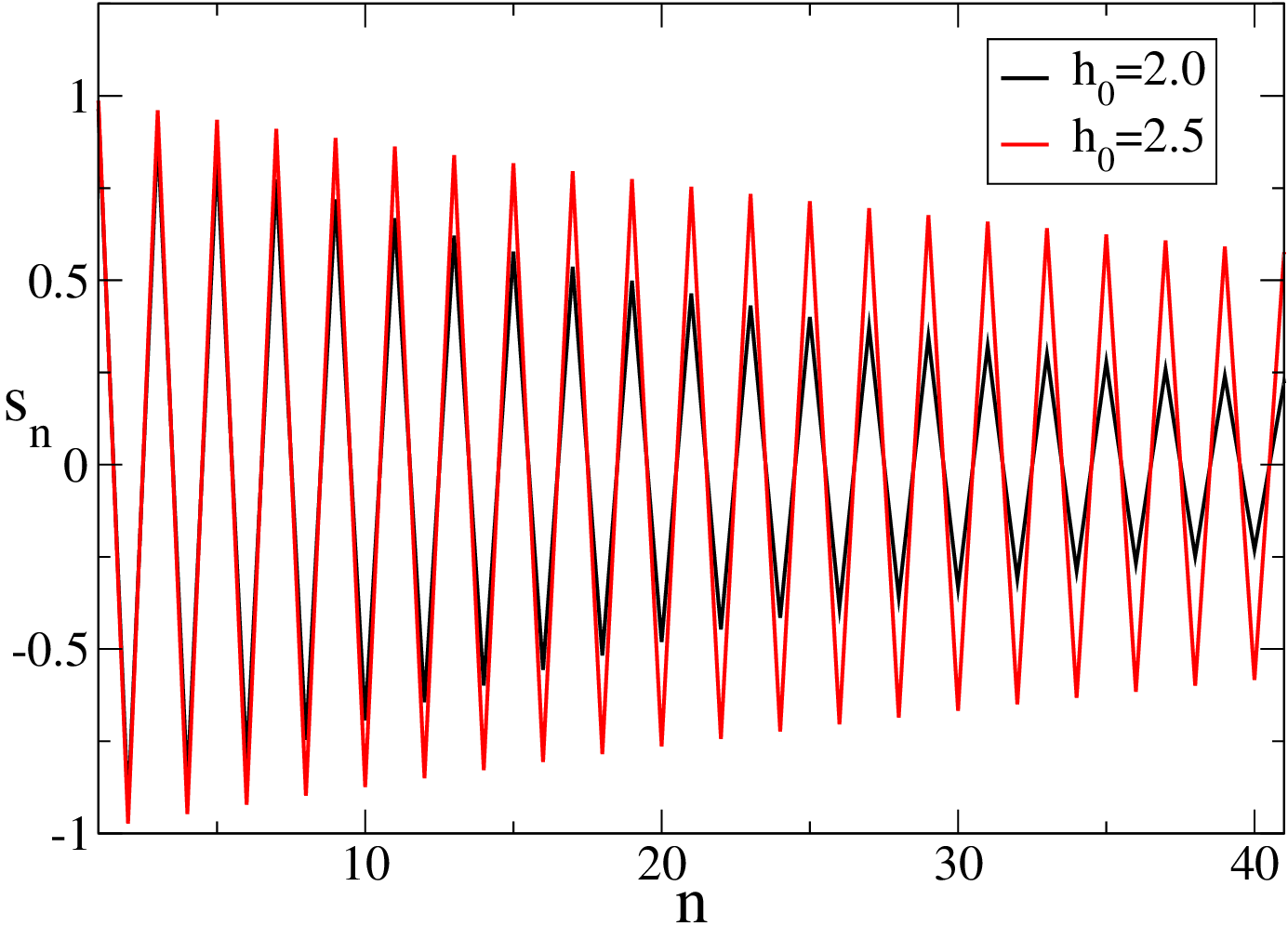}
\vspace{-2mm}
\caption{Oscillations in the two-state model. The inverse temperature is 
set to $\beta=1$. The average spin orientation $s_n$ is given in Eq. \eqref{eqavgs}}
\label{fig1} 
\end{figure}

We assume that $\tau$ is large as compared to the relaxation time to reach the stationary distribution. 
The probability to be in state $s=1$ at the end of the nth time-interval $p_n$ is then
\begin{equation}
p_n= \frac{\textrm{e}^{\beta h_n}}{2\cosh(\beta h_n)}.
\end{equation}

From the feedback rule that the the external field for the next time-interval  has the opposite 
sign to the orientation of the spin at the end of the present time-interval, we obtain that the probability $p_n$ follows the recursion relation 
\begin{equation}
p_{n+1}= (1-p_n)p+p_n(1-p),
\label{eqrecursion}
\end{equation}
where 
\begin{equation}
p= \frac{\textrm{e}^{\beta h_0}}{2\cosh(\beta h_0)}.
\label{eqdefp}
\end{equation}
As initial condition we set $h_1=h_0$ for the first time-interval $n=1$. The solution of Eq. \eqref{eqrecursion} is 
given by     
\begin{equation}
p_n= \frac{1}{2}\left[1-(1-2p)^n\right].
\end{equation}
Hence, the average spin orientation $s_n=p_n(+1)+(1-p_n)(-1)$ reads
\begin{equation}
s_n=  (-1)^{n+1}[\tanh(\beta h_0)]^n,
\label{eqavgs}
\end{equation}
where we have used Eq. \eqref{eqdefp}. As shown in Fig. \ref{fig1}, $s_n$ oscillates between positive and negative values with a period $n_{\textrm{osc}}=2$ in terms 
of the integer $n$. In terms of time such oscillations correspond to a period $2\tau$, where $\tau$ is the time-interval between two measurements. Oscillations in 
feedback driven systems with a discrete feedback scheme are sub-harmonic, i.e., they have a period of oscillation larger than the natural period of the feedback driven system $\tau$.

\subsection{Relation between precision and work}

The amplitude of the oscillations decay exponentially since $\tanh(\beta h_0)\le 1$. This damping of the oscillations 
in the average spin orientation is related to noise. If we consider two different stochastic trajectories,  after some time 
they will have different phases due to fluctuations. The number of coherent oscillations that characterizes the precision of the oscillations is 
defined as the ratio of the decay time and the period of oscillation. If we rewrite Eq. \eqref{eqavgs} as 
\begin{align}
s_n & =  \cos(\pi n+\pi)\textrm{e}^{-n[-\ln \tanh(\beta h_0)]}\nonumber\\
 &\equiv \cos(2\pi n/n_{\textrm{osc}}+\pi)\textrm{e}^{-n/n_{\textrm{dec}}},
\label{eqavgs2}
\end{align}
we obtain the decay time $n_{\textrm{dec}}=\{-\ln[ \tanh(\beta h_0)]\}^{-1}$. The number of coherent oscillations is then
\begin{equation}
\mathcal{N}\equiv \frac{n_{\textrm{dec}}}{n_{\textrm{osc}}}=  \frac{\{-\ln[ \tanh(\beta h_0)]\}^{-1}}{2} .
\label{eqNcohe}
\end{equation}

Even though the transition rates during a time-interval fulfill detailed balance,
the feedback procedure drives the system out of equilibrium. The average 
work exerted on the system per time-interval is 
\begin{equation}
W_{\textrm{avg}}= p(2h_0)-(1-p)(2h_0)=2h_0\tanh(\beta h_0).
\label{eqW}
\end{equation}
There are two contributions to the work per period. One is the probability to finish
a time-interval with the spin and the field pointing in the same direction $p$ multiplied 
by the energy difference between the two states $2h_0$. The other is the probability to finish
a time-interval with the spin and the field pointing in different directions $1-p$ multiplied 
by the energy difference between the two states $2h_0$. Unlike the spin orientation, this average 
work does not oscillate. It is stationary already after the first time-interval. 

\begin{figure}
{\psfrag{ln(N)}{$\ln(\mathcal{N})$}\includegraphics[width=77mm]{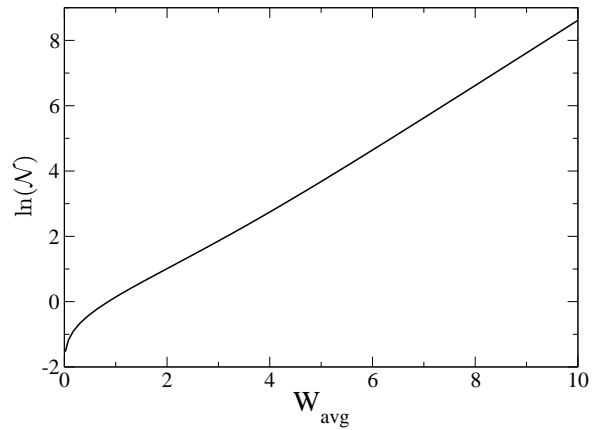}
\vspace{-2mm}
\caption{Parametric plot of  $\ln(\mathcal{N})$ versus $W_{\textrm{avg}}$ for the two-state model.
The inverse temperature is $\beta=1$ and the external field is varied from $h_0=0.1$ to 
$h_0=5$.}
\label{fig2}} 
\end{figure}

The relation between precision, as characterized by $\mathcal{N}$, and energy consumption 
that is quantified by $W_{\textrm{avg}}$ is analyzed in Fig. \ref{fig2}, where we plot $\mathcal{N}$ as 
a function of $W_{\textrm{avg}}$. First, the number of coherent oscillations increases with an increasing 
energy consumption. Second, at equilibrium ($h_0=0$) there are no oscillations. The same 
property is true for oscillations in autonomous systems such as biochemical oscillators,  since
energy dissipation is a  general necessary condition for the onset of oscillations. Third, in the limit $\beta h_0\to\infty$,
both the number of coherent oscillations $\mathcal{N}$ and the the work exerted on the system 
$W$ diverges. This property is in stark contrast with coherent oscillations in autonomous 
systems. For this case, even in a limit of divergent energy dissipation the number 
of coherent oscillations is finite and essentially bounded by the number of states \cite{bara17a}. 
This fundamental difference between oscillations in feedback driven systems  and autonomous systems is a main result.    
The possibility of an indefinite number of coherent oscillations in a finite system in the presence of thermal fluctuations is not exclusive to feedback driven systems. 
Subharmonic oscillations in periodically driven systems also show this property \cite{ober19}. A relevant difference between these two cases is that the minimal model for 
a periodically driven system analyzed in \cite{ober19} has three states, whereas our minimal model has two states.

\subsection{Informational thermodynamic cost}

A distinctive feature of a feedback driven system is that the thermodynamic cost is not only quantified by the the work 
$W$ but also by the mutual information term $I$, as we report  in Appendix \ref{appa}.
For this model the information obtained by the measurements is given by 
\begin{equation}
I= -p\ln p-(1-p)\ln (1-p).
\label{eqI}
\end{equation}
Since there is no measurement error,
this quantity is just the Shannon entropy of a two-states system  at the 
end of a time-interval, where $p$ is the probability of the lower energy state.  This quantity is stationary  given that $p$ is the same 
at the end of all time-intervals.
The informational thermodynamic cost $I$, as compared to the work $W_{\textrm{avg}}$, has a different 
relation with the number of coherent oscillations. For $\beta h_0\to\infty$, which leads to indefinite oscillations, this cost is minimal, i.e., this limit leads to $p=1$, which 
leads to $I=0$. For $h_0=0$, for which there are no oscillations,  the mutual information is maximal $I=\ln 2$. As the number of coherent oscillations $\mathcal{N}$ increases, by increasing the parameter $h_0$, the informational thermodynamic cost $I$ decreases. This term  plays a key role in the entropy balance of more complex feedback driven system as we will show in the following section.

\section{Ising model with feedback}
\label{sec3}
\subsection{Model definition}

We now consider a fully connected Ising model with $N$ spins and a total of 
$2^N$ states. The energy of the system is 
\begin{equation}
E^{h}_M= -J M^2/(2N)-hM,
\label{eqenergy}
\end{equation}
where the magnetization takes the values $M=-N,-N+2,\ldots,N-2,N$, $J$ is the coupling parameter, 
and $h$ is the external field. The state of the system is fully characterized by the orientation of 
all the $N$ spins. However, since the energy of the mean-field model only depends on the magnetization 
$M$, the dynamics during a time-interval can be simplified to a random walk on the $M$ space with transition 
rates fulfilling the detailed balance condition. In particular, we choose 
the transition rates  
\begin{equation}
w^{h_n}_{M\to M+2}\equiv \frac{\gamma (N-M)\textrm{e}^{\beta[J(m+N^{-1})+h_n]}}{2\cosh(\beta[J(m+N^{-1})+h_n])}
\label{eqrates2}
\end{equation}
and
\begin{equation}
w^{h_n}_{M\to M-2}\equiv \frac{\gamma (N+M)\textrm{e}^{-\beta[J(m-N^{-1})+h_n]}}{2\cosh(\beta[J(m-N^{-1})+h_n])},
\label{eqrates3}
\end{equation}
where the subscript $n$ in $h_n$ represents the nth time-interval and $\gamma$ is a parameter that sets the time-scale of the transition rates. The duration of the time interval is $\tau$. We point out that 
these transition rates depend on the measurement outcome in the previous time-interval and, therefore, they 
depend on the particular stochastic trajectory.

The feedback scheme is as follows. At the end of a time-interval, the state of the 
system is measured with perfect precision and the magnetic field $h$ is changed according to     
\begin{equation}
h_{n+1}= h_n-\alpha M_n/N\equiv h_n-\alpha m_n,
\label{eqfeedback}
\end{equation}
where $\alpha$ is a constant and $M_n$ is the magnetization at the end of the nth time-interval. There is a similarity between this feedback scheme and the feedback scheme 
for the two-state model. If the average magnetization is negative then the minimum of the  free energy is 
on the negative side. The feedback is such that the minimum of the free energy is shifted towards the positive 
side due to the change in the external field. For the opposite case of a positive magnetization,  
the feedback scheme changes the minimum towards the negative side. Hence, it is expected that this feedback scheme
generates oscillatory behavior.

Numerical simulations of this model were performed as follows. We use the Gillespie algorithm \cite{gill77} to simulate a continuous 
time random walk with the rates given by Eq. \eqref{eqrates2} and Eq. \eqref{eqrates3} for a time-interval $\tau$. At the end of the time-interval 
the transition rates are updated by a change in the magnetic field given by Eq. \eqref{eqfeedback}. The initial condition for our 
simulations was $h=0$ for the external field and $M=N$ for the magnetization. 

\subsection{Oscillations in the Ising model}

\begin{figure}
\includegraphics[width=77mm]{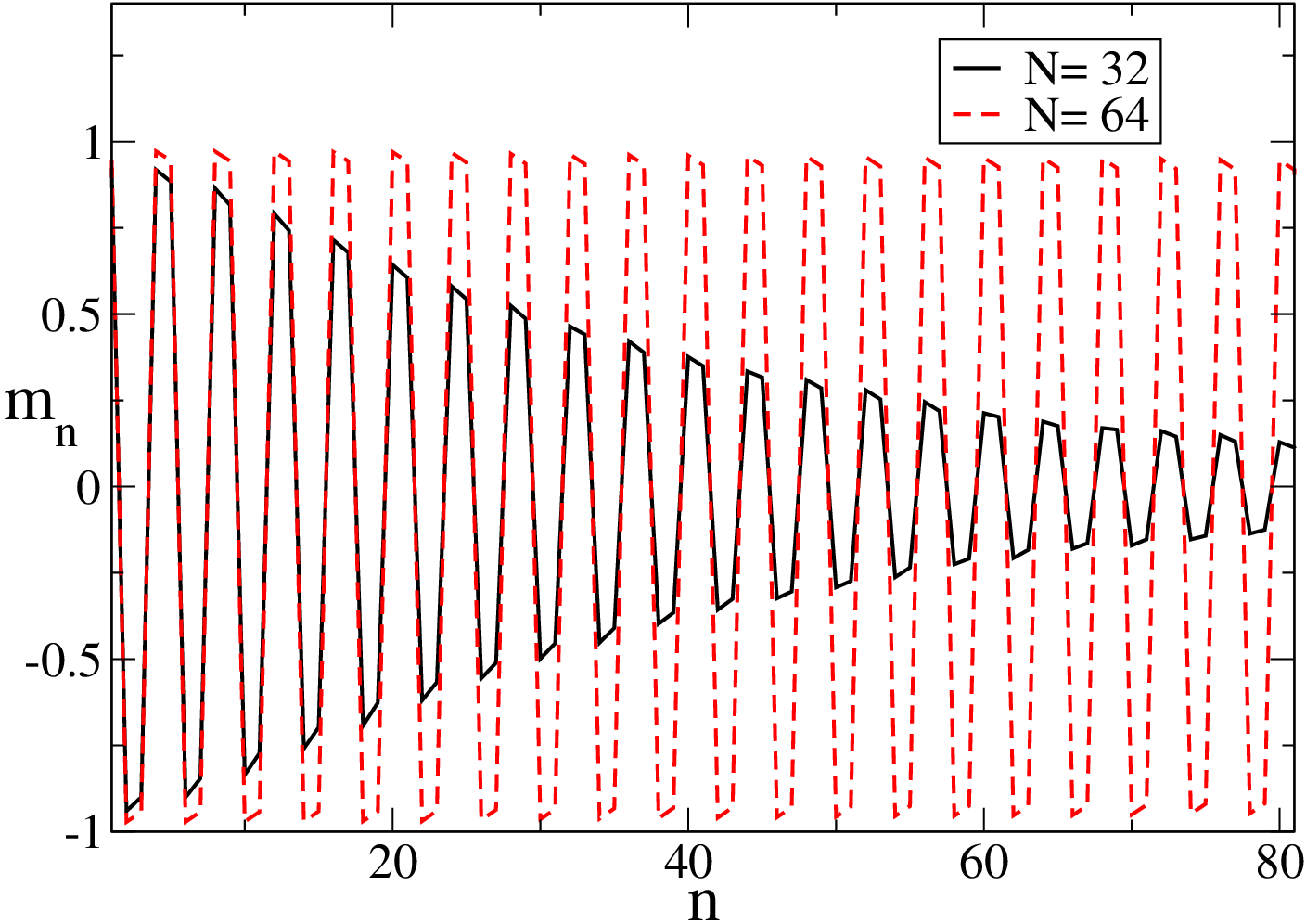}
\vspace{-2mm}
\caption{Oscillations in the feedback driven Ising model. Average magnetization $m_n$
as a function of $n$. The parameters are set 
to $\gamma=1$, $\tau=100$, $J=1$, $\beta= 2$, and $\alpha=0.5$. The period of 
oscillations, which is the same for both system size for this value of $\alpha$,
is $n_{\textrm{osc}}=4$.}
\label{fig3} 
\end{figure}

This feedback driven Ising model displays oscillations in the magnetization  $m_n$ for temperatures below the 
critical temperature ($T_c=J^{-1}$). As shown in Fig. \ref{fig3}, the number of coherent oscillations depends on the 
system size  and becomes indefinite  in the thermodynamic limit.  
This feature has been demonstrated analytically for a model with continuous feedback  \cite{dema19}.

The oscillatory behavior of the magnetization shown in Fig. \ref{fig3} is similar to sub-harmonic oscillations in periodically-driven 
systems with many degrees of freedom, such  as time-crystals. An example related to our model is the 
periodically driven Ising model analyzed in \cite{gamb19b}, which displays subharmonic oscillations with a period that is two times the natural 
period of the drive. For the oscillations in our model with the parameters used in Fig \ref{fig3}, the period is four times the 
time-interval between two measurements.

\begin{figure}
\includegraphics[width=77mm]{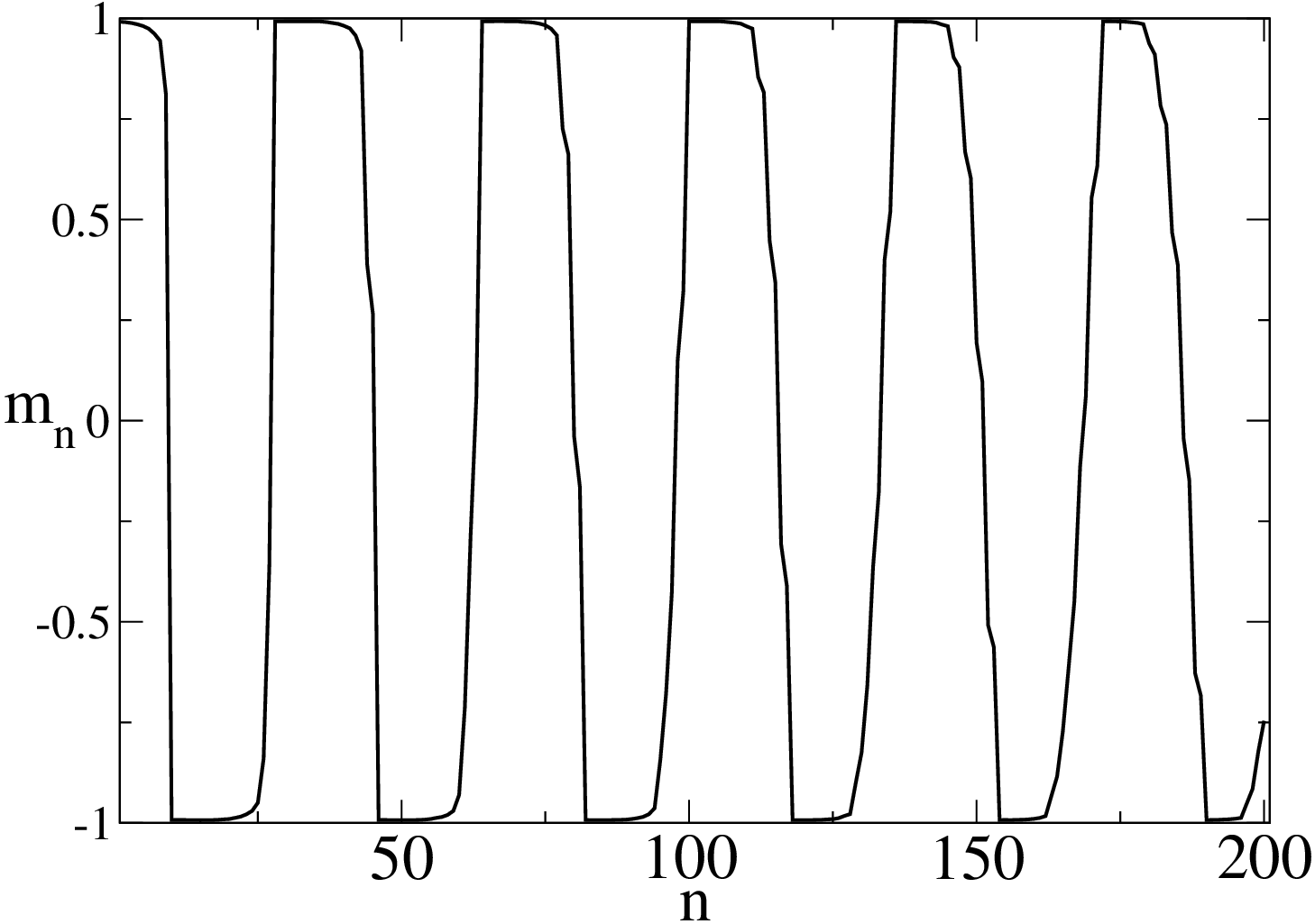}
\vspace{-2mm}
\caption{Effect of the parameter $\alpha$ in the period of oscillation. Average magnetization $m_n$
as a function of $n$. The parameters are set 
to $\gamma=1$, $\tau=100$, $J=1$, $\beta= 3$, $\alpha=0.05$, and $N=256$. The period of 
oscillations is estimated to be $n_{\textrm{osc}}= 36$.}
\label{fig4} 
\end{figure}

The period of oscillations has a strong dependence on the parameter $\alpha$. In Fig. \ref{fig4} we show that the period of oscillation becomes 
much larger for $\alpha=0.05$, as compared to the oscillations shown in Fig. \ref{fig3} with $\alpha=0.5$. For this case of a smaller 
$\alpha$ the period of oscillations is estimated to be $n_{\textrm{osc}}= 36$. Besides the parameter $\alpha$, numerical simulations show that the period depends also on the inverse temperature $\beta$.

\subsection{Work and heat for the Ising model}

Let us consider a stochastic trajectory of the fully connected Ising model. We denote 
by $M_n$ the magnetization at the end of the nth time-interval. From Eq. \eqref{eqworkgen}
in Appendix \ref{appa} the total work per spin exerted on the system for a stochastic trajectory with $\nu$ time intervals is   
\begin{equation}
W=\frac{1}{N}\sum_{n=1}^{\nu-1}\left(E_{M_{n}}^{h_{n+1}}-E_{M_{n}}^{h_n}\right),
\label{eqworkmain}
\end{equation}
where $E_{M_{n}}^{h_{n+1}}$ is the energy of the system in the beginning of the (n+1)th time-interval and $E_{M_{n}}^{h_n}$
is the energy of the system at the end of the nth time-interval. We point out that we do not carry out the explicit dependence 
of the work $W$ on the stochastic trajectory as we do in Appendix \ref{appa}. Furthermore, the quantity $W$ in Appendix \ref{appa}
represents the total work whereas here it represents the work per spin, i.e., the work divided by the number of spins $N$. This 
quantity is finite in the thermodynamic limit $N\to\infty$. For all thermodynamic quantities of the Ising model, such as heat and entropy change,
we consider the thermodynamic quantity per spin, hence, there is a factor $N^{-1}$ in relation to the generic expressions given in Appendix \ref{appa}.
From Eqs. \eqref{eqenergy} and \eqref{eqfeedback}, the work in Eq. \eqref{eqworkmain} becomes
\begin{align}
W & =\sum_{n=1}^{\nu-1}(h^{n}-h^{n+1})M_n/N= \sum_{n=1}^{\nu-1} \alpha (M_n)^2/N^2\nonumber\\
& \equiv \sum_{n=1}^{\nu-1}W_n.
\label{eqworkmain2}
\end{align}
The dissipated heat per spin is obtained from Eq. \eqref{eqheatgen} in Appendix \ref{appa}, together with Eqs. \eqref{eqenergy} and \eqref{eqfeedback},    
\begin{align}
Q &=\sum_{n=1}^{\nu}\left(E_{M_n}^{h_n}-E_{M_{n-1}}^{h_n}\right)\nonumber\\
    &= \sum_{n=1}^{\nu}J(M_n^2-M_{n-1}^2)/(2N^2)+h^n(M_n-M_{n-1})/N\nonumber\\
    &\equiv \sum_{n=1}^{\nu-1}Q_n.
\end{align}
The quantity $W_n$ is the work exerted on the system at the end of period $n$ and $Q_n$ is the heat dissipated during the nth time-interval. There is an abuse of notation 
to represent the averages of $M_n$, $W_n$, and $Q_n$, which are stochastic quantities. In all figures and in all expressions below these symbols 
represent averages over stochastic trajectories.   

\begin{figure}
\includegraphics[width=77mm]{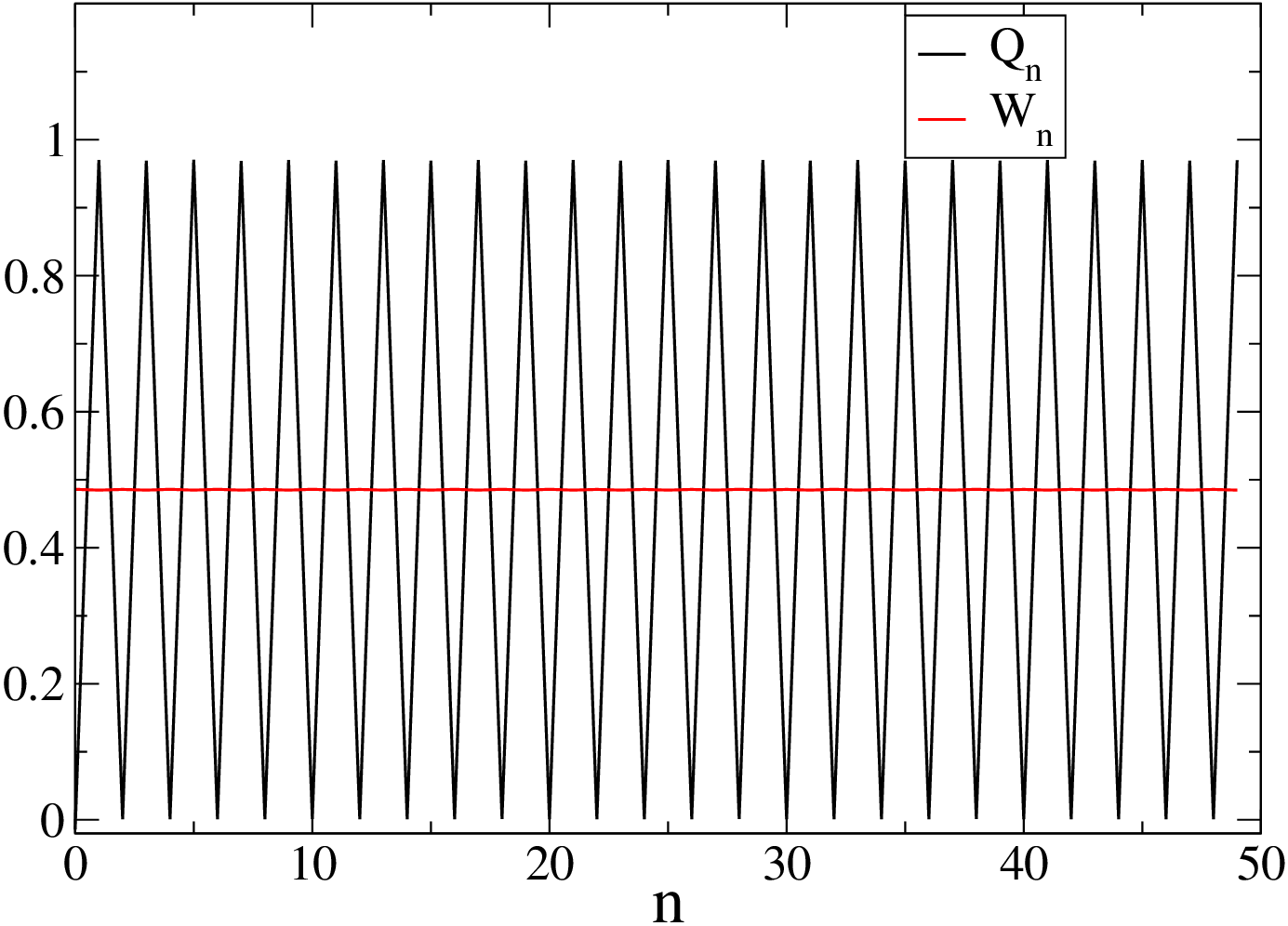}
\vspace{-2mm}
\caption{Oscillations in work and heat . Average work $W_n$ and 
average heat $Q_n$ as a function of $n$. The parameters are set 
to $\gamma=1$, $\tau=100$, $J=1$, $\beta= 3$, $\alpha=0.5$, and $N=128$. The amplitude 
of oscillations for the work $W_n$, which cannot be seem in this resolution, are much smaller than the amplitude of oscillations for the 
heat $Q_n$.}
\label{fig5} 
\end{figure}

In Fig. \ref{fig5} we plot heat $Q_n$ and work $W_n$ as a function of $n$. Both quantities oscillate with a period that 
is half of the period of oscillations of the magnetization (which is $n_{\textrm{osc}}=4$ for this case), 
since they are both quadratic functions of the variables $m$ and $h$. The amplitude of the oscillations for the work 
$W_n$ are much smaller than the amplitude of the oscillations in the heat $Q_n$.

\subsection{Second law and information}

\begin{figure}
\includegraphics[width=77mm]{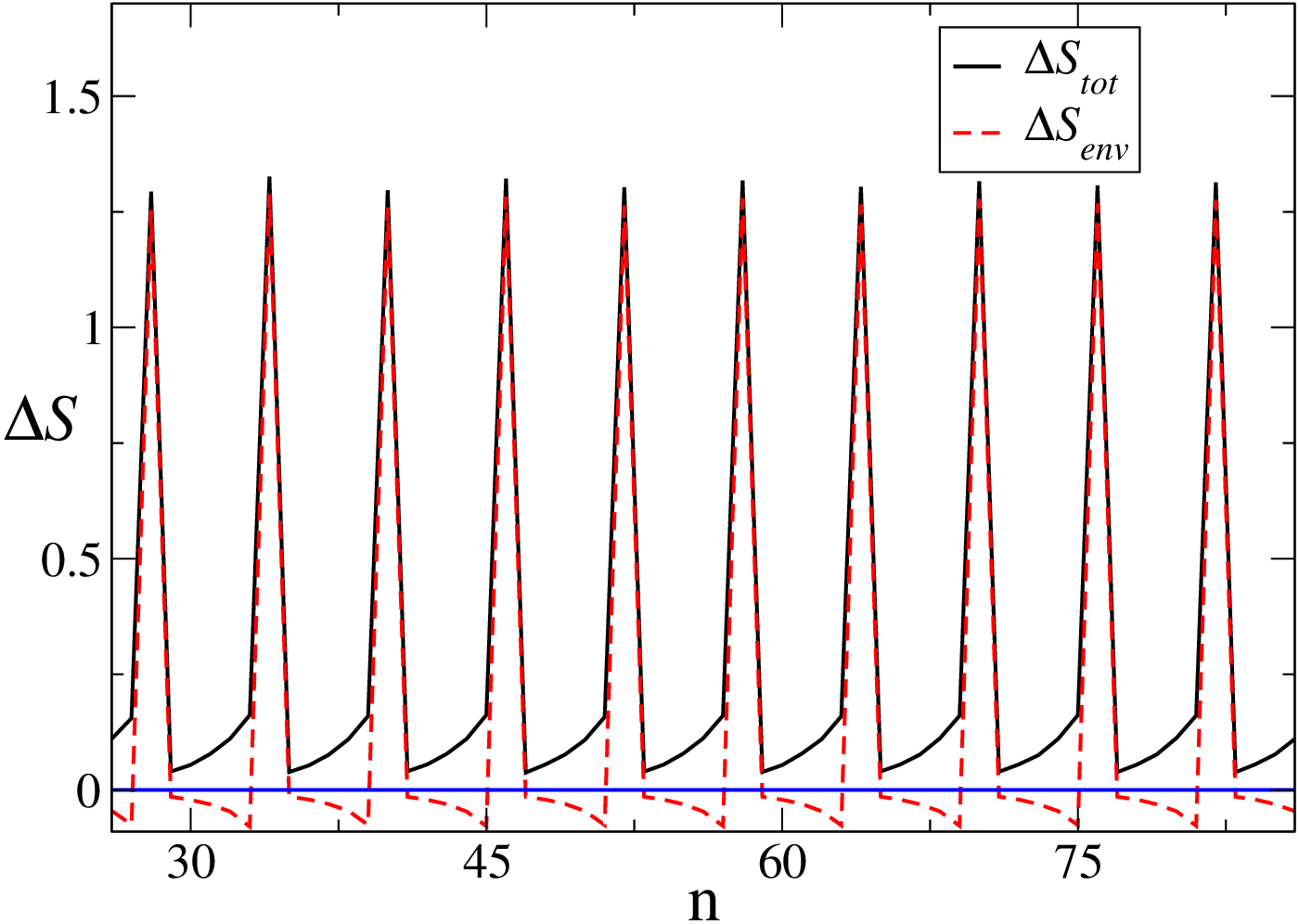}
\vspace{-2mm}
\caption{Second law for the feedback driven Ising model. The entropy change of the environment $\Delta S_{\textrm{env}}^n$, which can be negative, and the 
total entropy change $\Delta S_{\textrm{tot}}^n=\Delta S_{\textrm{env}}^n+I_n$ as functions of $n$. The parameters are set 
to $\gamma=1$, $\tau=100$, $J=1$, $\beta= 2$, $\alpha=0.1$, and $N=512$.}
\label{fig6} 
\end{figure}

We now analyze the second law for the Ising model with feedback. The average entropy increase of 
the external environment per spin for the nth time interval is given by $\Delta S_{\textrm{env}}^n= \beta Q_n$.
As shown in Fig. \ref{fig6}, this oscillating quantity can be negative at certain times $n$. 
Such negative dissipated heat for a system in contact with a single heat bath would constitute a ``violation'' of the standard statement of 
the second law of thermodynamics for systems without feedback. 
However, for feedback driven systems there is also the informational contribution contribution 
$I_n$. The total entropy change per spin for the nth time interval is $\Delta S_{\textrm{tot}}^n=\Delta S_{\textrm{env}}^n+I_n\ge 0$, 
where this second law inequality is discussed in Appendix \ref{appa}. In Fig. \ref{fig6} we show that while the entropy change of the environment for a 
certain times $n$ can be negative, when we also account for the informational term the total entropy change $\Delta S_{\textrm{tot}}^n$ is positive, 
as predicted by the second law for feedback driven systems.

The mutual information $I_n$ was calculated in the following way. Since there are no measurement errors the mutual information $I_n$ is just the entropy of the system at 
the the end of the nth time-interval. If we denote a spin configuration with $N$ spins by $\mathbf{s}$ 
then the mutual information per spin is  
\begin{equation}
I_n= -\frac{1}{N}\sum_\mathbf{s} P_n(\mathbf{s})\ln [P_n(\mathbf{s})],
\label{eqIns}
\end{equation}
where $P_n(\mathbf{s})$ is the probability of the spin configuration $\mathbf{s}$ at the end of the nth time interval.
The sum in Eq. \eqref{eqIns} is over the $2^N$ spin configurations.
For the present mean field model with the Hamiltonian in \eqref{eqenergy} that only depends on the magnetization 
$M$, we have 
\begin{eqnarray}
P_n(\mathbf{s})= P(\mathbf{s}|M) P_n(M) \nonumber\\
P(\mathbf{s}|M) = \frac{\delta(\sum_i s_i,M) }{\mathcal{C}_{N,M}},
\label{eqPnM}
\end{eqnarray}
where $P_n(M)$ is the probability of magnetization $M$ at the end of the nth time interval and $P(\mathbf{s}|M)$ is the conditional probability of the spin configuration given the magnetization $M$, 
which is uniform over the  spin configurations with magnetization $M$. The number of spins configurations with magnetization $M$ is  
\begin{equation}
\mathcal{C}_{N,M}=\frac{N!}{((N-M)/2)!((N+M)/2)!}.
\end{equation}
From Eq. \eqref{eqIns} and Eq. \eqref{eqPnM} we obtain 
\begin{equation}
I_n=-\frac{1}{N}\sum_M P_n(M)\ln(P_n(M))+\frac{1}{N}\sum_M P_n(M)\ln(\mathcal{C}_{N,M}).
\end{equation}
The sum in this equation is over the $N+1$ possible values of the magnetization.
The mutual information $I_n$ can then be evaluated from a numerical calculation 
of the probability $P_n(M)$.

\subsection{Scaling of the entropy production}

\begin{figure} 
  \centering
  \includegraphics[width=77mm]{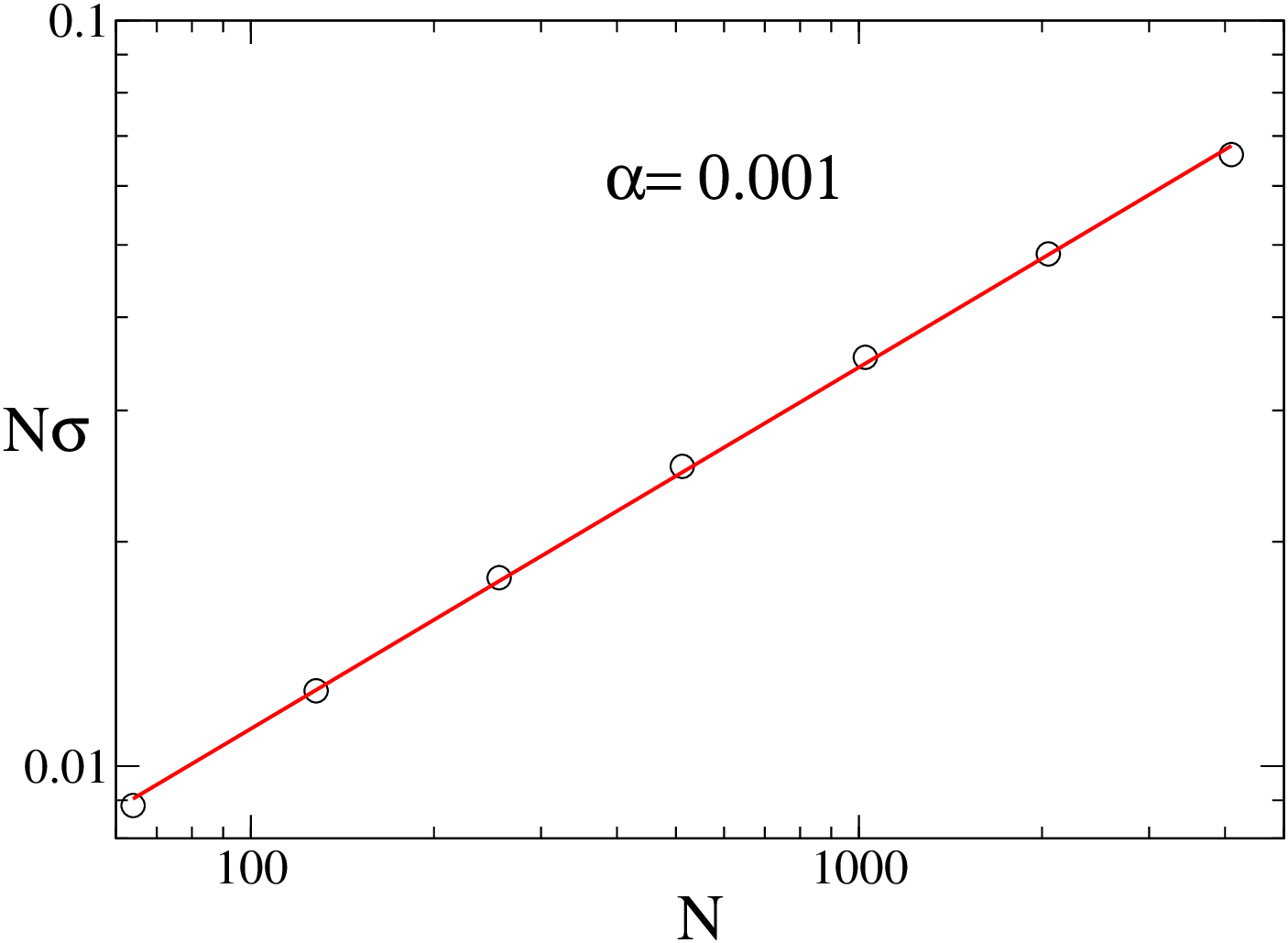}
  \includegraphics[width=77mm]{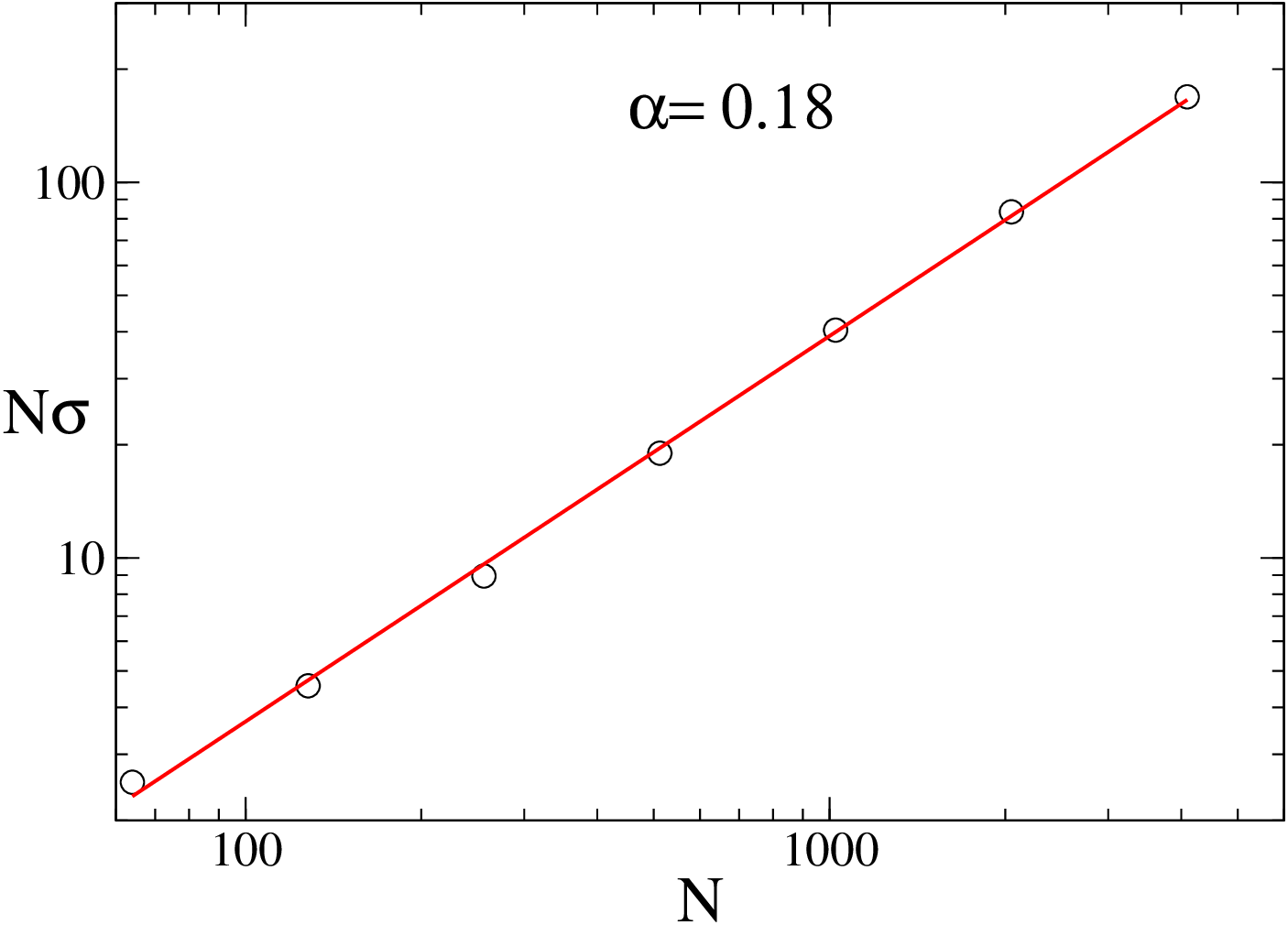}
  \includegraphics[width=77mm]{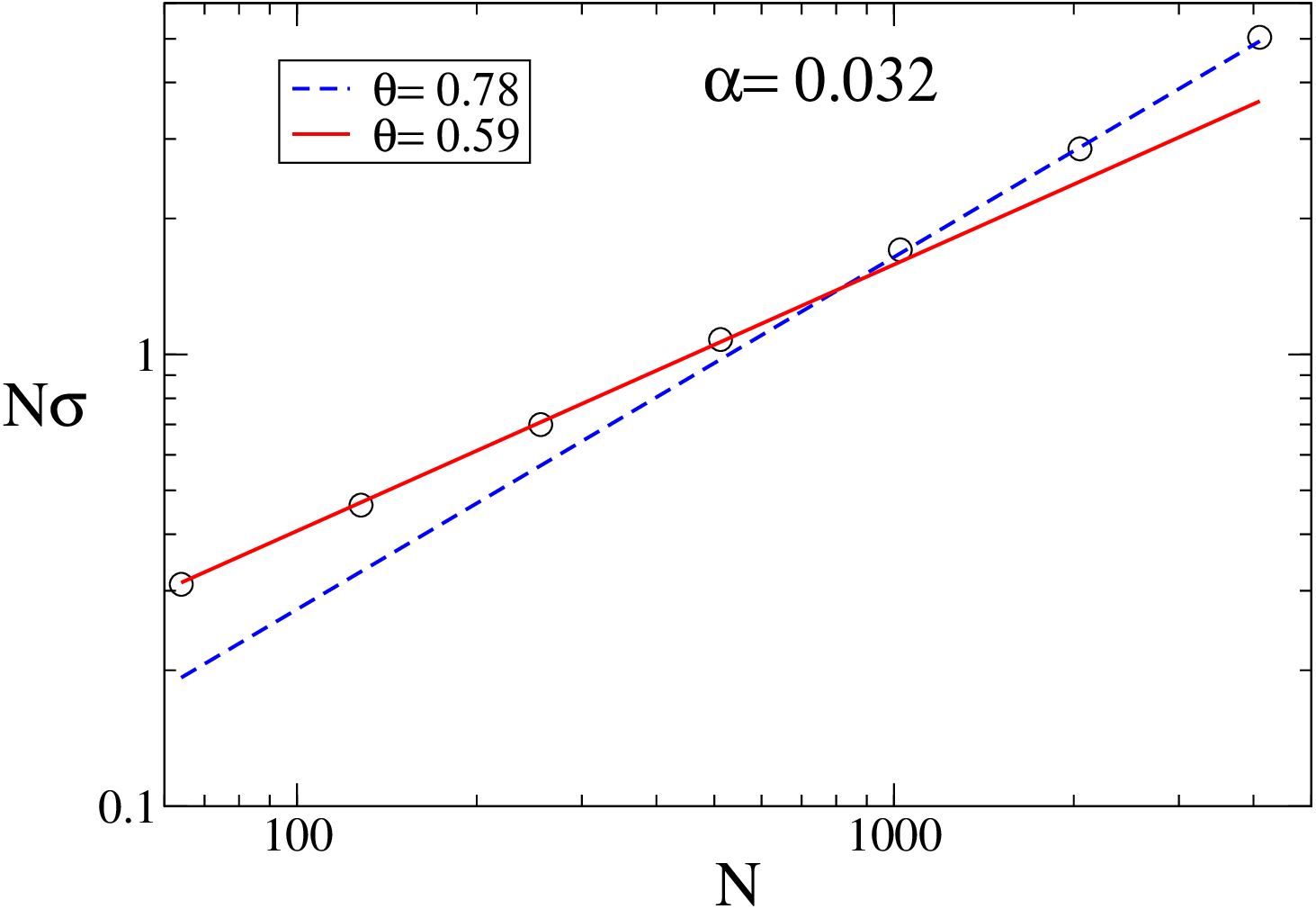}
  \caption{The entropy production rate $\sigma$ as a function of the number of spins $N$ at the critical point $\beta=1$ for different values of $\alpha$. 
  For the upper panel that is associated with $\alpha= 0.001$, the exponent estimated with the full red line is compatible with $\theta=0.5$. For the mid panel 
  that is associated with $\alpha=0.18$, the exponent estimated with the full red line is compatible with $\theta=1$. For the lower panel that is associated with $\alpha=0.032$ 
  the effective exponent increases for increasing $N$. Parameters are set to $\gamma=1$, $\tau=100$, and $J=1$.}
\label{fig7}
\end{figure}

The stationary average change of entropy production per spin and per time-interval is defined as 
\begin{equation}
\sigma= \frac{\beta}{\nu}\lim_{\nu\to \infty} \sum_{n=1}^\nu Q_n. 
\end{equation}
We have analyzed numerically the scaling behavior of this quantity of the number of spins $N$.  Above the critical 
temperature  $\sigma$ tends to a constant value in the thermodynamic limit. Below the critical point 
$\sigma$  goes to zero in the thermodynamic limit. At the critical point this quantity shows a scaling 
behavior that depends on the parameters $\alpha$ and $\tau$. We define the exponent $\theta$ as 
\begin{equation}
\sigma\sim N^{1-\theta}. 
\end{equation}
In Fig. \ref{fig7}, we plot $N\sigma$ as a function of $N$ for different values of $\alpha$ at fixed $\tau$. 
For smaller values of $\alpha$ we obtain an exponent compatible with the mean field value of a continuous transition $\theta=0.5$.

For larger values of $\alpha$ we obtain an exponent compatible with $\theta=1$. 
Hence, the entropy production is finite in the thermodynamic limit 
at the critical point, i.e., for larger values of $\alpha$ there is a first-order phase transition. For intermediate values of $\alpha$ we obtain an effective exponent 
between $0.5$ and $1$. However, the effective exponent estimated within a larger $N$ region 
is larger than the effective exponent for smaller values of $N$. Hence, for intermediate values of $\alpha$ there is a transient in $N$ which 
goes beyond the values of $N$ used in our simulations. 
In Appendix \ref{appb} we characterize analytically the scaling behavior and the phase transition for small $\alpha$ and $\tau$ in the model with continuous feedback.

\section{Conclusion}
\label{sec4}

We have provided a systematic analysis of oscillations in noisy feedback driven systems.
The two-state model introduced here provides arguably the simplest example of such oscillation. 
Exact calculations with this simple model  demonstrate fundamental differences
between oscillations in feedback driven systems and the other two kind of oscillators. 
Importantly, even in a two-state system with thermal fluctuations the oscillations can remain 
coherent for an arbitrarily long time, in contrast to oscillations in autonomous systems, 
which can only remain coherent for a finite time that is determined by the number 
of states of the system \cite{bara17a}. This property of indefinite oscillations in finite 
systems is also present for subharmonic oscillations in periodically driven systems \cite{ober19},
however, the minimal model in this case was found to have three states, whereas our minimal 
of a feedback driven oscillator has two states.

The feedback driven fully connected Ising model provides an example of a system with 
several degrees of freedom that has oscillations that become indefinite in the thermodynamic limit, as previously demonstrated 
in \cite{dema19} for a model with continuous feedback. We have shown that the model analyzed here with discrete feedback 
displays subharmonic oscillations in the magnetization similar to the subharmonic oscillations in time-crystals. 

The thermodynamic cost of oscillatory feedback driven systems has to be carefully analyzed. We have shown that 
the oscillatory dissipated heat for the Ising model is negative at certain times, even if the system is 
in contact with a single heat bath. However, if the informational thermodynamic cost related to measurements is 
also taken into account the oscillatory total entropy change per time-interval is positive at all times, as predicted 
by the second law for feedback driven systems. Whereas negative dissipated heat in feedback driven systems is  a 
well known fact, previous studies have not considered the second law for feedback driven systems with oscillatons.       

A similarity between oscillations in feedback driven systems and in autonomous systems deserves to be mentioned. 
In both cases, in the deterministic limit, 
we have instances of self-oscillatiors \cite{andr66} that cannot be differentiated at this level of description. However, these feedback driven oscillators
and autonomous oscillators are two different classes of oscillators at the stochastic level of description. In particular, 
they have different limitations concerning the precision of oscillations and the second law of thermodynamics implies different 
inequalities, with an informational term showing up for feedback driven oscillators.

Concerning future work, it would be interesting to numerically analyze the critical behavior of a two-dimensional Ising model with a feedback 
scheme similar to the one considered here. In particular, a comparison of such model with  time-crystals could lead to an understanding 
of differences and similarities between oscillatory feedback driven systems and time-crystals, concerning their critical behavior and thermodynamics.       
From a broader perspective, a theoretical framework for oscillations in stochastic systems and their relation to thermodynamics is emerging. 
The application of this framework to understand biological oscillators and to produce optimal synthetic oscillators remain key open problems.

\appendix

\section{Feedback driven systems}
\label{appa}

\subsection{Definition}

In this appendix we define thermodynamic quantities, such as heat, work, and entropy, and 
discuss the second law for feedback driven systems. A more general theory that includes 
a fluctuation theorem for feedback driven systems can be found in \cite{saga12}. Such systems are characterized 
by Markovian dynamics during time-intervals of duration $\tau$ and by measurement 
and feedback at the end of each time-interval. Mathematically, feedback translates into  
transition rates for the present time-interval that depend on the measurement outcome 
at the end of the previous time-interval. Transition rates are then random variables 
that depend on the particular stochastic trajectory.

We consider a discrete set of states of the system $x=1,2,\ldots,\Omega$.
A stochastic trajectory with total time $\nu\tau$ is denoted by $X_\nu$, where $\tau$ is 
the duration of time-interval. The state of the system and the measurement outcome at the end of the nth time-interval are 
denoted by $x_n$ and $y_n$, respectively. The stochastic trajectory $Z_{\nu\tau}$ can be written as 
\begin{equation}
Z_{\nu}= (X^{\lambda_1}_{\tau},y_1,X^{\lambda_2}_{\tau},y_2,\ldots,X^{\lambda_{\nu-1}}_{\tau},y_{\nu-1},X^{\lambda_{\nu}}_{\tau}).
\end{equation}
The variable $\lambda_n$ represents  the protocol during the nth interval. This protocol  $\lambda_n$  depends on the measurement outcome 
at the end of the previous time-interval $y_{n-1}$. Here we restrict to the case of time-independent protocols during the time-interval $\tau$.      
The stochastic trajectory $Z_\nu$ is a sequence of sub-trajectories $X^{\lambda_n}_{\tau}$ and measurement outcomes 
$y_n$. Each sub-trajectory is Markovian and can be written as $X^{\lambda_n}_{\tau}=(x_i^n,x_1^n,\ldots,x_f^n)$, where $x_f^n=x_i^{n+1}=x_n$. 
For convenience  we write this sub-trajectory as discrete in time. The number of elements in the trajectory is the 
inverse of the time-step multiplied by $\tau$. If we take this time-step to go 
to zero we recover the continuous-time description.

The measurement outcome $y_n$ is obtained with a preassigned conditional probability $P(y_n|x_n)$. Here, we assume that 
the measurement outcome is independent of the measurement history  and only depends on the state of the system $x_n$.
The number of possible states for the measurement outcome can be smaller than $\Omega$, which is the number of 
states of the system. For example, for the Ising model analyzed in Sec. \ref{sec3}, the number of states 
is $\Omega=2^N$, whereas the magnetization that is the outcome of the measurement has $N+1$ possible states.

The transition rate at the nth time-interval from state $x$ to state $x'$ is denoted by $w^{\lambda_n}_{xx'}$.
Here, we assume that the transition rates during a time-interval are time-independent and fulfill the detailed balance 
relation with some energy function $E_x^{\lambda_n}$, i.e., 
\begin{equation}
\frac{w^{\lambda_n}_{xx'}}{w^{\lambda_n}_{x'x}}= \textrm{e}^{\beta(E_x^{\lambda_n}-E_{x'}^{\lambda_n})}.
\label{eqgeneralDB}
\end{equation}

\subsection{Work and heat}

The stochastic work is defined as 
\begin{equation}
W[Z_\nu]=\sum_{n=1}^{\nu-1}(E_{x_n}^{n+1}-E_{x_n}^n).
\label{eqworkgen}
\end{equation}
The work exerted on the system is the sum of the changes in energy at the end of a time-interval
due to the feedback scheme. As an example, the change in the magnetic field due to feedback 
for the models analyzed here lead to a change in the energy of the system.

Each jump in the sub-trajectory $X^n_{\nu}$ changes the entropy of the environment, which is connected with the 
dissipated heat. In particular, the entropy change of the environment for a jump 
that changes the state of the system from $x$ to $x'$ is $\Delta S_{\textrm{env}}=\ln(w_{xx'}/w_{x'x})$. 
This formula can be seen as a postulate of stochastic thermodynamics \cite{seif12}. From Eq. \ref{eqgeneralDB}, 
we obtain that the entropy change of the external environment associated with the whole sub-trajectory 
$X^n_{\nu}$ as $\beta(E_{x_{n-1}}^n-E_{x_{n}}^n)$. The entropy change associated with the trajectory $Z_\nu$
is then  
\begin{equation}
\Delta S_{\textrm{env}}[Z_\nu]=\sum_{n=1}^{\nu}\beta(E_{x_{n-1}}^n-E_{x_{n}}^n).
\end{equation}

The dissipated heat $Q[Z_\nu]= \frac{1}{\beta}\Delta S_{\textrm{env}}[Z_\nu]$, is then
given by   
\begin{equation}
Q[Z_\nu]=\sum_{n=1}^{\nu}(E_{x_{n-1}}^n-E_{x_{n}}^n).
\label{eqheatgen}
\end{equation}
From Eq. \eqref{eqworkgen} and \eqref{eqheatgen} we obtain the first law of thermodynamics 
\begin{equation}
\Delta E[Z_\nu]=W[Z_\nu]-Q[Z_\nu]=E_{x_{\nu}}^{\nu}-E_{x_{0}}^1,
\label{eqfirstlaw}
\end{equation}
where $\Delta E[Z_\nu]$ is the energy change associated with the trajectory $Z_\nu$.

\subsection{Second law}
 
The total entropy change in a feedback driven system is composed by the change of the 
entropy of the external environment, the change of the entropy of the system, and 
change of entropy associated with the information obtained with the measurements. We now 
consider average entropy changes, instead of the stochastic quantities from the previous subsection.
An average here means an average over all possible stochastic trajectories.
The average entropy change of the external environment is denoted by   $\Delta S_{\textrm{env}}$.

Each measurement at the end of time-interval reduces the uncertainty about the state of the system. 
In the case of perfect measurements that we consider in the models analyzed here, the uncertainty 
about the state of the system is completely eliminated. This reduction of uncertainty is accompanied by 
a reduction of the entropy of the system. Such reduction of entropy must be compensated by an increase of 
entropy in somewhere else. In other words, a controller that makes measurements and apply feedback according 
to the measurement outcomes imply an increase of entropy  \cite{cao09}. 

At the end of the nth time-interval, and an instant before the measurement is take, the average entropy of the system is
$H^n(x)\equiv -\sum_x P^n(x)\ln P^n(x)$, where $P^n(x)$ is the probability to be in state $x$
at the end of nth time-interval. After the measurement the entropy is reduced to  
$H^n(x|y)= -\sum_{x,y}P^n(x,y)\ln P^n(x|y)$. This entropy can be calculated with the knowledge of 
$P^n(x)$ and $P(y|x)$, the preassigned conditional probability of the measurement outcome that 
is independent of $n$. The joint probability is given by $P^n(x,y)=P^n(x)P(y|x)$. From Bayes' theorem
the conditional probability $P^n(x|y)$ is $P^n(x|y)= P^n(x)P(y|x)/P^n(y)$, where $P^n(y)=\sum_xP^n(x,y)$.

The total entropy increase to compensate for the entropy reduction of the system after a measurement from $H^n(x)$ to $H^n(x|y)$ 
is the mutual information  
\begin{equation}
I^n\equiv H^n(x)-H^n(x|y).
\end{equation} 
This mutual information quantifies the minimal entropy increase that a controller acting on the feedback driven system generates. 
Mutual information is a standard quantity in information theory and it has the property $I^n\ge 0$ \cite{cove06}. The action of 
a controller cannot decrease the total entropy. For the case of the models analyzed here with perfect measurements $H^n(x|y)=0$,
leading to $I^n= H^n(x)$.  We point out that for more general feedback driven systems with a feedback scheme 
that can depend on the measurement history the informational observable that quantifies the entropy change due to the action of 
the controller is the transfer entropy \cite{saga12}.

The informational change in entropy due to the action of a controller is then
\begin{equation}
\Delta S_{\textrm{inf}}=\sum_{n=1}^{\nu-1} I^n.
\end{equation} 
Furthermore, the change in the entropy of the system is  
\begin{equation}
\Delta S_{\textrm{sys}}= H^\nu(x)-H^0(x).
\end{equation} 
An important difference between $\Delta S_{\textrm{sys}}$ and the other two entropy changes is that $\Delta S_{\textrm{env}}$ and $\Delta S_{\textrm{inf}}$ are both extensive in the number of time intervals $\nu$, whereas 
$\Delta S_{\textrm{sys}}$ does not increase with an increase in  $\nu$.

Finally, the second law for feedback driven systems reads 
\begin{equation}
\Delta S_{\textrm{tot}}= \Delta S_{\textrm{env}}+\Delta S_{\textrm{inf}}+\Delta S_{\textrm{sys}}\ge 0.
\end{equation}

For this second law we have considered a total time-interval $\nu\tau$. However, this second law is valid for any 
total time-interval. In particular let us take a total time interval that starts after the measurement at the end of the $(n-1)$th
time-interval and finishes after the measurement at the end of the nth time-interval. Furthermore, we also assume perfect measurements, which reduce the entropy 
of the system to $0$. Hence, $\Delta S_{\textrm{sys}}=0$. The second law for such total time-interval then reads  
\begin{equation}
\Delta S_{\textrm{tot}}^n= \Delta S_{\textrm{env}}^n+I_n\ge 0,
\end{equation}
where $\Delta S_{\textrm{env}}^n$ is the average entropy change of the environment associated with the nth time-interval. This inequality is 
the one illustrated in Fig. \ref{fig6}.

\section{Model with continuous feedback}
\label{appb}

\subsection{Langevin equation and thermodynamic observables}

We now consider the Ising system with continuous feedback \cite{dema19}, i.e., a feedback scheme that is applied to the system at every instant.
The phenomenological Langevin equation for this Ising model reads  
\begin{eqnarray}
dm &=& \{ -m +\tanh[\beta(m+h)]\}dt + b dw \nonumber\\
dh &=& -c m dt 
\label{eqlangevin}
\end{eqnarray}
where the coupling parameter of the Ising model is set to $J=1$, the magnetization $m$ is a continuous variable between $-1$ and $1$, $w$ represents the Wiener process, and the noise strength is 
\begin{equation}
b =  \sqrt{\frac{2}{\beta N}}.
\label{eqnoise}
\end{equation}
The feedback scheme represented by the second equation in Eq. \eqref{eqlangevin} is equivalent to the feedback scheme represented by Eq. \eqref{eqfeedback} for the model with discrete feedback in the limit of 
$\tau$ and $\alpha$ very small such that their ratio is finite and given by
\begin{equation}
c=\alpha/\tau.
\end{equation}
In the thermodynamic limit, for which the noise term in Eq. \eqref{eqlangevin} is negligible,   the phase transition with spontaneous symmetry breaking in the standard Ising model without feedback  is substituted by an Andronov-Hopf bifurcation in this model with feedback, with the onset of oscillations
below the critical point \cite{dema19}.   
The  energy per spin of the Ising model is $u = -\frac{1}{2} m^2 -hm$, the infinitesimal change in energy then reads  
\begin{equation}
du= -(m+h) dm -m dh - b^2 dt,    
\end{equation}
where we have used Ito's differentiation rule. The infinitesimal work per spin exerted on the system due to the change in the external field $h$ is     
\begin{equation}
dW= -mdh = c m^2 dt. 
\label{eqdefwcont} 
\end{equation}
This expression is equivalent to the expression in Eq. \eqref{eqworkmain2} for the model with discrete feedback. From the first law, we obtain the infinitesimal  dissipated heat per spin as 
\begin{equation}
dQ=  dW -du 
\label{eqfirstlaw2}
\end{equation}
These three differentials are stochastic quantities. The average rate of entropy production is defined as  
\begin{equation}
\sigma\equiv \lim_{T \to \infty } \frac{1}{T} \int^T \langle \frac{dQ}{dt} \rangle dt,  
\label{eqsigmacont}
\end{equation}
where the brackets denote an average over stochastic trajectories. Note that $\lim_{T \to \infty } \frac{1}{T} \int^T \langle \frac{du}{dt} \rangle dt=0$
since the energy difference $u(T)-u(0)$ is not extensive in $T$. Hence, from the first law in Eq. \eqref{eqfirstlaw2}, we obtain 
\begin{equation}
\sigma = \lim_{T \to \infty } \frac{1}{T} \int^T \langle \frac{dW}{dt} \rangle dt= \lim_{T \to \infty } \frac{1}{T} \int^T c\langle m^2 \rangle dt,  
\label{eqsigmacont2}
\end{equation}
where the second equality follows from Eq. \eqref{eqdefwcont}.

\subsection{Analytical calculations for the rate of entropy production}

\begin{figure}
\includegraphics[width=77mm]{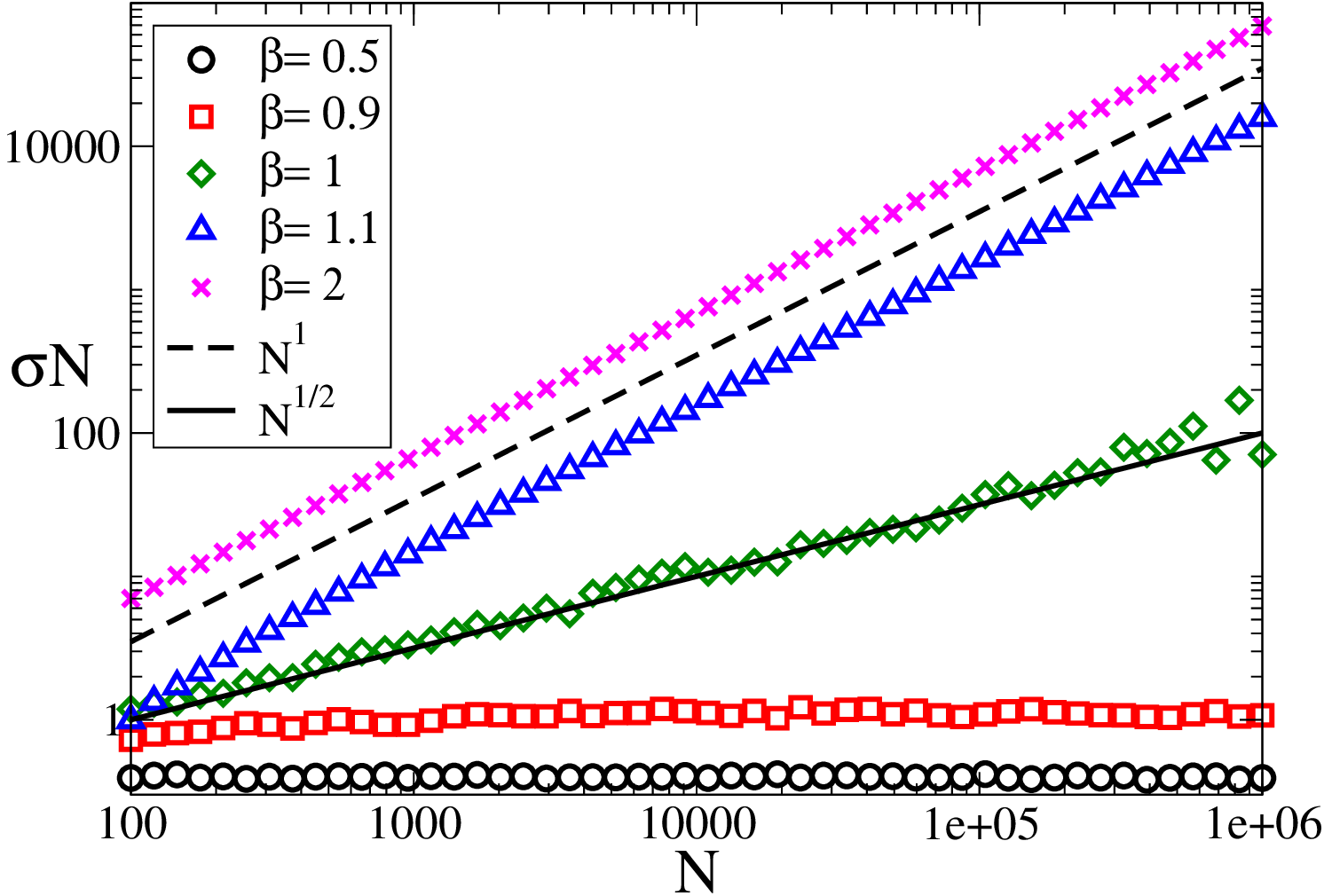}
\vspace{-2mm}
\caption{Scaling of the average rate of entropy production $\sigma N$ in the fully connected Ising model with feedback as a function of the system size $N$ for several temperatures above, at and below the critical point $\beta=2,1.1,1,0.9,0.5$ (and $c=0.1$), from numerical simulations.}
\label{figapp} 
\end{figure}

For this model, we derive with an  analytical argument, the following scaling law for the average rate of entropy production,
\begin{equation}
\sigma=
\begin{cases}
\mathcal{O}(1/N) \quad \beta<1\\
\mathcal{O}(1/\sqrt{N}) \quad \beta=1\\
\mathcal{O}(1) \quad \beta>1  
\end{cases}
\end{equation}
This analytical calculation is in agreement with numerical simulations shown in Fig. \ref{figapp}.   

First, we consider the case $\beta <1$. Upon linearizing Eq. \eqref{eqlangevin} around the stationary point $(m^*=0,h^*=0)$,
we obtain
\begin{eqnarray}
dm &=& ( (\beta-1)m +\beta h )dt + b dw \\
dh &=& - c m  dt
\end{eqnarray}
The stationary value for the square of the magnetization is  $\langle m^2 \rangle_s = b^2/2$. Hence, from Eq. \eqref{eqnoise} and from Eq. \eqref{eqsigmacont}, we obtain that $\sigma= c \beta^{-1} N^{-1}$.

Second, we consider the case  $\beta  \gtrsim 1$. In the thermodynamic limit the dynamics of the system can be mapped into the equation for the Van der Pol oscillator \cite{dema19}
\begin{equation}
\ddot{m} + (\beta-1-m^2)\dot{m} +\sqrt{c} m=0.
\end{equation}
Performing an expansion in $\beta-1$, the solution reads
\begin{eqnarray}
m(t) \sim 2 \sqrt{\beta-1} \cos(\sqrt{c} t) + \mathcal{O}(\beta-1) \nonumber \\
h(t) \sim -2 \sqrt{c(\beta-1)} \sin(\sqrt{c} t) + \mathcal{O}(\beta-1) 
\end{eqnarray}
The system is performing harmonic oscillations with the conserved quantity
\begin{equation}
E\equiv m^2 + h^2 /c = 2 (\beta-1) 
\label{eqdefE}
\end{equation}
From expression \eqref{eqsigmacont} we obtain that  the average rate of entropy production is $\sigma=2(\beta-1)$.

Third, we consider the model at the critical point $\beta=1$.  
An expansion of $\tanh[\beta(m+h)]$ in Eq. \eqref{eqlangevin} leads to 
\begin{eqnarray}
dm &=& (h -m^3/3)dt + b dw \nonumber\\
dh &=& -c m  dt
\label{eqdefE2}
\end{eqnarray}
Upon considering the quantity  $E$ defined in Eq. \eqref{eqdefE}, together with Eq. \eqref{eqdefE2}, we obtain the following SDE,
\begin{equation}
dE = (-2/3 m^4 +b^2)dt +2 m dw.
\end{equation}
Since $E$ is bounded, the time derivative of its average becomes zero in the steady state, which  implies the scaling  $\langle m^4 \rangle_s = 3b^2/2$. This last equation implies the square root scaling 
for the average entropy production per spin with the system size $N$ at the critical point.



\bibliographystyle{apsrev4-1}

\bibliography{refs} 

\end{document}